\documentclass{cimento}

\definecolor{slateblue}{rgb}{0.2,0.2,0.6}
\usepackage[colorlinks=true, linkcolor=slateblue, citecolor=slateblue, urlcolor=slateblue, hyperindex=true, bookmarks=true, implicit=false]{hyperref}

\newcommand{\NTIN}[5]{\textit{#1}, \href{http://dx.doi.org/#5}{\textbf{#2} (#3) #4}}

\usepackage{graphicx} 

\title{Development of a web application for monitoring solar activity and cosmic radiation}
\shorttitle{A web application for monitoring solar activity and cosmic radiation}
\author{D.~Pelosi\,\from{ins:x}\ETC,
N.~Tomassetti\,\from{ins:x},
M.~Duranti\,\from{ins:y}
}

\instlist{
  \inst{ins:x} Dipartimento di Fisica e Geologia, Universit\`a degli Studi di Perugia, Italy
  \inst{ins:y} INFN - Sezione di Perugia - Perugia, Italy
  }

\begin{document}
\maketitle

\begin{abstract}
The flux of cosmic rays (CRs) in the heliosphere is subjected to remarkable time variations caused by the 11-year cycle of solar activity. To help the study of this effect, we have developed a web application (Heliophysics Virtual Observatory) that collects real-time data on solar activity, interplanetary plasma, and charged radiation from several space missions or observatories. As we will show, our application can be used to visualize, manipulate, and download updated data on sunspots, heliospheric magnetic fields, solar wind, and neutron monitors counting rates. 
Data and calculations are automatically updated on daily basis. 
A nowcasting for the energy spectrum of CR protons near-Earth is also provided using calculations and real-time neutron monitor data as input. 
\end{abstract}

\section{Introduction}
During their motion inside the heliosphere, cosmic rays (CRs) experience the effects of heliospheric forces, commonly known as solar modulation. Specifically, the solar wind and its embedded magnetic field constantly reshape their energy spectra. 
The solar modulation effect is caused by several processes such as convection, drift motion, diffusion, and adiabatic cooling, although investigations are underway on defining the associated parameters \cite{ref:Potgieter2013}. As a result, the energy spectrum of CRs observed near-Earth is significantly different from that in the surrounding interstellar medium, the so-called Local Interstellar Spectrum (LIS). 
Furthermore, solar modulation is known to be energy- and time-dependent. In fact, the effect is more evident for CRs with kinetic energies below $\sim$\,10\,GeV and shows a clear correlation with solar activity. This implies that solar modulation inherits the quasi-periodical behavior of solar activity. More specifically, the monthly SunSpot Number (SSN) observed on the Sun's photosphere, widely used as a good proxy for solar activity, varies with a period of 11 years, known as solar cycle. 
A deep investigation of the solar modulation phenomenon is of crucial importance 
to achieve a full understanding of the dynamics of charged particles in the heliospheric turbulence, as well as to  accurately predict the radiation dose received by electronics and astronauts. 
Forecasting the CR fluxes near-Earth and in the interplanetary space is essential, 
given the ever-growing number of satellites orbiting around Earth and human space missions to Moon and Mars planned in the next decades. For this purpose, several analytic and numerical models of solar modulation have been 
proposed\,\cite{ref:Potgieter2013,ref:FF,ref:NTPRD,ref:NTAPJ}.
Recent progress in this field has been possible thanks to time-resolved data on CRs fluxes released from many space missions such as EPHIN/SOHO (since 1995 to 2018) \cite{ref:EPHINSOHO}, PAMELA (2006-2016) \cite{ref:PAMELA}, AMS-02 (since 2011 and still operative for all ISS lifetime) \cite{ref:AMS}, 
along with the direct LIS data from the Voyager probes in the interstellar space \cite{ref:VOYAGER1}.
The temporal variations of CRs fluxes are also measured, with some caveats, with the ground-based network of Neutron Monitors (NMs) whose data are collected in since 1951 \cite{ref:NM}.
A NM detector is an energy-integrating device whose count rate ($N$) is defined (for each species of CR) as an integral, above the local geomagnetic rigidity cutoff, of a product of the near-Earth CR flux and the specific yield function of the detector.
Models also need solar data such as SSN (in different time resolutions) provided by the SILSO/SIDC database of the \textit{Royal Observatory of Belgium}~\cite{ref:SILSO}, polar field strength and tilt angle values of the heliospheric current sheet (HCS) monthly provided by the \textit{Wilcox Solar Observatory}~\cite{ref:WILCOX} and heliospheric data about radial speed and proton density of solar wind, monthly distributed by NASA missions WIND and ACE~\cite{ref:NASA}. 
We have developed the \textit{Heliophysics Virtual Observatory} (HVO) in order to make data-access easier and faster. 
HVO is a web application that collects all the data mentioned above in a unique tool providing a daily automatic update. This tool gives users the functionalities of visualizing, manipulating and downloading updated data.
We also present a simplified real-time model of near-Earth proton flux integrated into a specific section of HVO.

\section{Real-time model}
The propagation of CRs in the heliosphere is governed by the Parker equation: 
\begin{equation}
\label{eq:1}
\frac{\partial f}{\partial t}  =  - (C \vec V + \langle \vec v_{drift} \rangle) \cdot \nabla f + \nabla \cdot ( \textbf{K} \cdot \nabla f) + \frac{1}{3} (\nabla \cdot \vec V)  \frac{\partial f}{\partial \ln R} + q
\end{equation}  
The equation describes the temporal evolution of CR phase space density $f=f(t,R)$, where $R=p/Z$ is the CR rigidity, $\langle \vec v_{drift} \rangle$ is the averaged particle drift velocity, $\vec V$ is the solar wind velocity, $\textbf{K}$ is the symmetric part of CR diffusion tensor, and $q$ is any local source of CRs \cite{ref:Potgieter2013}. 
The Parker equation is
often resolved within the so-called Force-Field (FF) approximation~\cite{ref:FF}. 
The FF model assumes steady-state conditions 
(i.e. negligible short-term modulation effects), 
radially expanding wind $V(r)$, isotropic 
and separable diffusion coefficient $K\equiv\kappa_{1}(r){\cdot}\kappa_{2}(R)$, 
negligible drift and loss terms. 
Despite these assumptions are often violated, the FF approximation provides a useful way to describe  
the near-Earth CR flux evolution and it is frequently used thanks to its simplicity.
The resulting CR flux $J(t,R)$ is related to $f$ by $J=R^{2}f$. Writing the solution in terms of kinetic energy per nucleon $E$, for a CR nucleus with charge number $Z$ and mass number $A$, the near-Earth ($r=1$\,AU) flux at the epoch $t$ is given by: 
\begin{equation}
\label{eq:2}
J(E,t) = \frac{(E + M_p)^2  - M_p^2}{(E + M_p +\frac{Z}{A}\phi(t)) ^2 -M_p^2}  J_{\rm{LIS}} (E+\frac{Z}{A}\phi(t)),
\end{equation}
where $\phi$ is the \emph{modulation potential}, it has the units of an electric potential, typically in the range 0.1-1 GV. The parameter $\phi$ can be interpreted as the averaged rigidity loss of CRs in their motion from the edge of the heliosphere down 
to the Earth.
Thus, the implementation of this simplified model depends on the knowledge of two key elements: the time-series of $\phi$ and the LIS.
In this work, we have used the new LIS models based on the latest results from Voyager 1 and AMS-02 \cite{ref:LIS} and the values of the modulation potential reconstructed by 
\textit{Usoskin et al.\,2011}\,\cite{ref:USOSKIN}, from NM data on monthly basis, since 1964 to 2011. 
To set up the $\phi$ reconstruction after 2011 and to the present epoch, instead of repeating the Usoskin methodology (based on NM's yield function), we propose a simplified method. 
For a given NM detector, the NM counting rate $N(t)$ and $\phi(t)$ are anti-correlated and we can establish a quadratic relation between them:
\begin{equation}
\label{eq:3}
\phi(N(t)) = A + B \cdot N(t) + C \cdot N(t)^2
\end{equation}
We determined the coefficient $A$, $B$, and $C$ as best-fit values using for several NM stations the Usoskin $\phi$-values. This enable us to obtain an prediction of $\phi$ for any epoch $t$ for which the NM rate $N$ is known.
Inserting  the parameterization of LIS and the $\phi$ value at epoch $t$ in Eq.\,\ref{eq:2} we obtain a real-time estimator for near-Earth CR flux. This simplified model has been integrated into HVO.

\section{The Heliophysics Virtual Observatory}
\begin{figure}[t]
\begin{center}
\includegraphics[width=0.89\columnwidth]{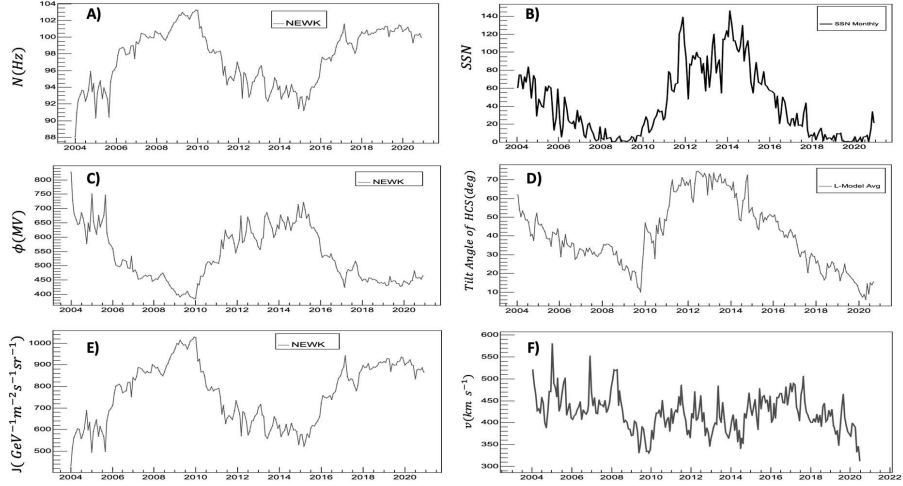}
\end{center}
\caption{%
Plots extracted from HVO:
A) Monthly rate $N(t)$ of the NEWK station (Newark, NJ, USA) in time interval 1/1/2004 - 1/1/2021.
B) Monthly SSN in the time interval 1/1/2004 - 1/1/2021.
C) Calculated time-series of $\phi$ using the NEWK rates (1/1/2004 - 1/1/2021).
D) Averaged tilt angle of the HCS measured with the classic model for any Carrington rotation between 1/1/2004 and 1/1/2021.
E) Estimated near-Earth proton flux at $E$ = 1 GeV using the NEWK rates (1/1/2004 - 1/1/2021). 
F) Solar wind speed for the period 1/1/2004 - 1/1/2021.
}
\label{fig1}
\end{figure}

The investigation of the solar modulation phenomenon requires a large variety of heliospheric and radiation data.
HVO \cite{ref:HVO} is a project developed 
under the CRISP scientific program of experimental study and phenomenological modeling of space weather, within the framework agreement between \textit{Università degli Studi di Perugia} and \textit{Agenzia Spaziale Italiana} (ASI).
HVO is a web application that daily extracts data with Python scripts from several databases (listed in Resources section), it visualizes and makes them available in a standardized format. HVO has been implemented with the JavaScript \texttt{ROOT} package \texttt{JSROOT}. It enables users to manipulate, directly from the web page, graphs and download data as text format or graphic objects with \texttt{ROOT} extension. To date, HVO has three main sections. The first one is dedicated to solar data such as SSN in daily, monthly, yearly, and smoothed formats extracted from SILSO/SIDC\,\cite{ref:SILSO}, observations of the Sun's polar magnetic field strength and tilt angle of the HCS reconstructed with the classic and radial model from the Wilcox Solar observatory~\cite{ref:WILCOX}. The second section contains heliospheric data such as proton density and radial speed of the solar wind, monthly updated from WIND and ACE~\cite{ref:NASA}. The third section contains cosmic radiation data from NMs and a real-time model for Galactic CR protons discussed in Sect.\,2. An interactive user interface provides the possibility to select one or more NM stations, choose the time resolution of the rates (daily, monthly, yearly and by Carrington rotation), set the proton energy and time interval.
HVO provides, for each selected NM, the graph of the count rate $N(t)$, the calculated time-series of $\phi$ from Eq.\,\ref{eq:3}, and the estimated  proton flux near-Earth $J(E)$ from Eq.\,\ref{eq:2}.
Fig.\,\ref{fig1} shows an example of the HVO functionality.

\section{Conclusions}
In this work, we have presented 
a web application aimed at monitoring solar activity and cosmic radiation, as well as providing real-time calculation of the energy spectra of CR protons in proximity of the Earth.
HVO is in its first development phase. We can propose possible improvements 
to realize a useful tool for the CR astrophysics and space physics community. 
In particular, we can extend the real-time proton model to other charged species. 
HVO can be also integrated with improved numerical models of CRs transport in the heliosphere~\cite{ref:NTPRD,ref:NTAPJ}, that will enable us to forecast the CR radiation at an interplanetary level. 
Finally, we can include 
other relevant observations such as, e.g., data on solar energetic particle (SEP) events, solar flares and coronal mass ejections (CME) or other interplanetary disturbance phenomena.
\acknowledgments
We acknowledge the support of Italian Space Agency under agreement ASI-UniPG 2019-2-HH.0.

\end{document}